\documentclass[prc,aps,twocolumn,groupedaddress,showpacs,floatfix]{revtex4}
\usepackage{graphicx}
\usepackage[usenames]{color}

\usepackage{amsbsy}

\begin{document}

\title{Phase separation in low-density neutron matter}

\author{ Alexandros Gezerlis$^{1}$ and Rishi Sharma$^2$ }
\affiliation{$^1$ Department of Physics, University of Washington, Seattle, Washington, 98195-1560, USA}
\affiliation{$^2$ Theory Group, TRIUMF, 4004 Wesbrook Mall, Vancouver, BC, V6T 2A3, Canada}

\begin{abstract}
Low-density neutron matter has been studied extensively for many decades, with
a view to better understanding the properties of neutron-star crusts and
neutron-rich nuclei.  Neutron matter is beyond experimental control, but in the past decade it
has become possible to create systems of fermionic ultracold atomic gases in a
regime close to low-density neutron matter.  In both these contexts pairing is
significant, making simple perturbative approaches impossible to apply and
necessitating \textit{ab initio} microscopic simulations. Atomic experiments
have also probed polarized matter. In this work, we study
population-imbalanced neutron matter (possibly relevant to magnetars and to
density functionals of nuclei) arriving at the lowest-energy configuration to
date.  For small to intermediate relative fractions the system turns out to be fully
normal, while beyond a critical polarization we find phase coexistence between
a partially polarized normal neutron gas and a balanced superfluid gas. As in
cold atoms, a homogeneous polarized superfluid is close to stability but not
stable with respect to phase separation. We also study the dependence of the
critical polarization on the density.
\end{abstract}

\date{\today}

\pacs{21.65.-f, 03.75.Ss, 05.30.Fk, 26.60.-c}

\maketitle

\section{Introduction}
\label{section:Introduction}

The phase structure of polarized (or population-imbalanced) Fermi gases is very rich. The
energetically favored phase depends on the masses and relative populations
of the species, the interaction strength, and whether the interactions are long
ranged. Polarized gases can exist in a variety of physical systems, including
ultracold trapped atomic Fermi gases~\cite{Giorgini:2008}, dense hadronic
phases near the cores of neutron
stars~\cite{Fantoni:2001,Vidana:2002,Rios:2005}, and possibly quark
matter~\cite{Alford:1999pa}.

In this paper we focus on the possibility that neutrons in the inner crust of
neutron stars are spin-polarized~\cite{Gezerlis:2011}.
Theoretical calculations suggest that this region, which is about a kilometer
thick, features unbound neutrons while the protons cluster in neutron rich
nuclei. While interactions with nuclei affect the properties of the unbound
neutrons~\cite{Pethick:2010zf,Cirigliano:2011}, in this work we consider an
isolated neutron fluid and compute the conditions for the coexistence of an
unpolarized phase and a polarized phase at densities roughly $30-250$ times
smaller than the nuclear saturation density. A polarized phase in the neutron
star crust will have a larger specific heat and therefore its presence may
affect the time scales associated with observed crustal transient
phenomena~\cite{Brown:2009}. Neutron superfluidity also affects thermal
transport properties~\cite{Aguilera:2009,Alford:2011}. If the polarization is
large enough all effects of superfluidity are expected to disappear. 

At the low densities we consider, the dominant term in the interaction between
spin-up and spin-down neutrons is an $s$-wave potential with a scattering length
$a\sim-18.6$ fm and range $r_e\sim2.7$ fm.  In a balanced system this attraction
drives the formation of Cooper pairs of spin-up and spin-down neutrons of
opposite momenta resulting in a neutron superfluid whose properties can be
understood qualitatively using BCS theory. All the neutrons are paired and the
number of up and down species is equal. Even at densities much smaller
than the nuclear saturation density, $a$ is much larger than the interparticle
spacing and obtaining quantitatively reliable results requires non-perturbative
techniques. Detailed Quantum Monte Carlo (QMC) simulations have calculated
the energy density ${\cal{E}}$ up to densities equal to or larger than the
nuclear saturation density~\cite{Carlson:Morales:2003,Gezerlis:2008}. At low
densities, the $s$-wave pairing gap obtained peaks at nearly $2$ MeV.

Magnetic fields tend to split the up and down Fermi surfaces thus disrupting the
pairing and creating a polarization. For magnetic fields ($B$) smaller than
$10^{16}-10^{17}$ Gauss, pairing wins (assuming no effect coming from the
lattice of nuclei), while larger fields can create a polarized phase. In BCS
theory, the critical field, or the resulting chemical potential split
($\delta\mu=g_N\mu_NB$ where $g_N\mu_N$ is the neutron magnetic moment), is
called the Clogston-Chandrasekhar point. At this point, the BCS superfluid can
coexist in equilibrium with a partially polarized normal phase which is
essentially a gas of free neutrons.  Numerically $\delta\mu_c=\Delta/\sqrt{2}$
where $\Delta$ is the gap in the BCS phase. A weakly polarized gas of neutrons
confined in a volume forms a mixed phase~\cite{Bedaque:2003} consisting of the
BCS and normal phases with volume fractions depending on the polarization.
Hartree corrections substantially modify the Clogston-Chandrasekhar point even
for modest couplings $|k_Fa|\sim0.5$, where $k_F=(3\pi^2
n)^{1/3}$ is the Fermi wave vector~\cite{Sharma:2008rc}. 

At strong coupling (mathematically given by $|k_Fa|\gtrsim1$) the energies of both the superfluid and the
normal phase need to be computed non-perturbatively. In this paper, we perform
these calculations for the normal phase and find the critical polarization at
which it can coexist with the superfluid phase. We also compare the energy of a
phase separated state with the previously evaluated~\cite{Gezerlis:2011}
homogeneous polarized superfluid phase. In this paper we only consider
competition between homogeneous and isotropic phases and ignore the possibility
of promising but still unconfirmed scenarios (like $p-$wave~\cite{Bulgac:2006} 
and inhomogeneous phases~\cite{Bulgac:2008}).

As mentioned above, our microscopic {\it ab initio} calculations may impact the observed 
behavior of neutron stars with extremely large internal magnetic fields. 
It is interesting to concurrently examine possible connections of such
simulations with terrestrial experiments. Our results may be useful for constraining Density
Functional Theories (DFTs) that are employed to study properties of heavy nuclei.
Equation of state results at densities close to the nuclear saturation density
have been used for some time to constrain Skyrme and other other density
functional approaches to heavy nuclei~\cite{Bender:2003}. 
The potential significance of such calculations has
led to a series of publications on the equation of state of low-density neutron
matter over the last few
decades~\cite{Friedman:1981,Akmal:1998,Carlson:Morales:2003,
Schwenk:2005,Margueron:2007b,Baldo:2008,Epelbaum:2008b,Epelbaum:2008a,
Abe:2009,Gandolfi:2009,Rios:2009,Hebeler:2010}. The density-dependence of the
$^1S_0$ gap in low-density neutron matter has also been used to constrain
Skyrme-Hartree-Fock-Bogoliubov treatments in their description of neutron-rich
nuclei~\cite{Chamel:2008}. 
Similarly, an ongoing project is studying the limit of extreme polarization in
neutron matter~\cite{Forbes:2011b}. This limit is known in condensed-matter physics as the
``polaron'' problem: one impurity embedded in a sea of fermions. Quantities of
interest here include the polaron binding energy and its effective
mass, which can be used to constrain Skyrme and other functionals.
In this line of study, DFTs can be constrained by
the equation of state of polarized unpaired neutron matter that we report here.

Another connection of our calculations with experiment may be possible through
the field of atomic ultracold Fermi gases. There, the scattering length of the
interaction between two hyperfine states near a Feshbach resonance can be
tuned by changing the magnetic field. Most experiments have been performed in
the region where the scattering length is infinite (unitarity limit) and the
effective range is
negligible~\cite{Luo:2009,Navon:2010,Carlson:2008,Schirotzek:2008,Shin:2008}.
QMC calculations performed with the same techniques that we use here in the
same limit give results that agree well with the experiments for both
unpolarized~\cite{Carlson:2003,Astrakharchik:2004,Gezerlis:2008} and
polarized~\cite{Carlson:2005,Lobo:2006,Pilati:2008,Gezerlis:2009} Fermi gases and this
gives us confidence in applying the same techniques to the problem with finite
$a$ and $r_e$. More directly, by tuning the scattering length and the effective
range~\cite{Marcelis:2008} it may be possible to obtain a system with similar
values of $k_F a$ and $k_F r_e$ as the inner crust and test our results.

\section{Construction of the mixed phase}
\label{section:MixedPhase}
The conditions for a polarized phase ($P$) to be in phase equilibrium with an
unpolarized superfluid ($SF$) phase are as follows:
\begin{enumerate}
\item{The pressures of the two phases should be equal.}
\item{The average chemical potential of the two species in the phases,
$\mu=(\mu_a+\mu_b)/2$, should be equal.}
\item{The chemical potential splitting, $\delta\mu=(\mu_a-\mu_b)/2$, should be 
less than the gap $\Delta$ in $SF$.}
\end{enumerate}
We construct the mixed phase between the superfluid and the partially polarized
normal phase ($NP$). The pressure in a polarized phase depends on the net
density of fermions ($n_a+n_b=n$) and their relative fraction $x=n_b/n_a$
where $a$ (or $\uparrow$) is the majority species. 

Motivated by the unitary Fermi gas~\cite{Cohen:2005ea,Bulgac:2007}, we write
down the energy density as: 
\begin{equation}
{\cal{E}} (n_a, n_b) = \frac{3}{5}\frac{(6\pi^2)^{2/3}}{2m}
(n_a g(x, n))^{5/3}\;.
\label{eq:gintro}
\end{equation}
For convenience we denote the energy per particle by $E$ and 
the corresponding energy for a free gas by $E_{FG}=3k_F^2/(10m)$. In terms of
$E$, $g(x, n) = (E/E_{FG(\uparrow)})^{3/5}$.

For the unitary gas, the discussion of mixed phases 
is simplified greatly since the function $g$ is only a function of $x$. In
neutron matter, the interaction potential has inherent scales (these can be 
thought of simplistically as $a$ and $r_e$). 
\begin{widetext}
To satisfy the conditions of phase
equilibrium, we need to calculate the chemical potentials and the pressure,
which include the functional dependence of $g$ on both $x$ and $n$.
The chemical potential is:
\begin{equation}
\mu = \frac{\partial {\cal{E}} }{\partial n} \Big |_{\delta n}\\
 = \frac{3^{2/3} \pi ^{4/3} \left(\frac{n }{x+1}\right)^{2/3}
   g(x,n)^{2/3} \left(\frac{2 n}{x+1}  \frac{\partial g}{\partial n}(x,n)+(1-x)
   \frac{\partial g}{\partial x}(x,n)+g(x,n)\right)}{2^{4/3} m}\;.
   ~\label{eq:mu}
\end{equation}

The chemical potential splitting is:
\begin{equation}
\delta\mu = \frac{\partial {\cal{E}} }{\partial \delta n} \Big |_{n}\\
=\frac{3^{2/3} \pi ^{4/3} \left(\frac{n }{x+1}\right)^{2/3}
   g(x,n)^{2/3} ((-x-1) \frac{\partial g}{\partial x}(x,n)+g(x,n))}
   {2^{4/3} m}\;.
   ~\label{eq:dmu}
\end{equation}

The pressure is simply $P = -{\cal{E}}(n, \delta n) + \mu n +
\delta\mu \delta n$, where $\delta n=n_a-n_b$.

For the unpolarized superfluid ($x=1$), the function $g$ depends only on the
density ($\delta n=0$) and we have:
\begin{equation}
\mu_{SF} 
 = \frac{3^{2/3} \pi ^{4/3} \left(\frac{n }{2}\right)^{2/3}
   g_{SF}(n)^{2/3} \left( n   \frac{\partial g_{SF}}{\partial n}(n)
   +g_{SF}(n)\right)}{2^{4/3} m}\;.
   ~\label{eq:muSF}
\end{equation}

\end{widetext}

To organize the discussion it is convenient to start from the completely
polarized ($x=0$) gas. Here the interactions between the only species present
are weak and the ground state is well described by a normal phase.  ($p-$wave
pairing and interactions will modify the energy of the fully polarized neutron
gas from a free Fermi gas. As we discuss in Sec.~\ref{section:MonteCarlo}, we
include the effect of $p-$wave interactions exactly, but ignore $p-$wave
pairing.) 

Now consider the following thought experiment. Keeping the density $n$
constant, change a few $a$ fermions to $b$ fermions so that $x$ changes.  For
small $x$ we expect that the phase obtained will be a partially polarized
normal phase. As we increase the fraction of $b$ fermions, it is possible that
some polarized superfluid phase ($SFP$) becomes the favored state of matter. 
Another possibility is that at some fractional density $x_c$, it becomes 
favorable to form a mixed phase between the partially polarized normal phase
($NP$) and the unpolarized superfluid phase ($SF$). In the present section, we will
focus on the mixed phase between $NP$ and $SF$. In
Sec.~\ref{section:MonteCarlo} we will calculate the energy for the polarized
normal gas ($NP$) phase and in
Sec.~\ref{section:PhaseCompetition} we will look at the competition 
between the homogeneous polarized superfluid ($SFP$) 
of Ref. \cite{Gezerlis:2011} and the mixed phase.

Equating the chemical potentials and the pressures gives two
equations in two variables, $x_c$ and $n_{SF}$. $n_{SF}$ is the total 
superfluid density and $n_{NP}$ is the total density of the normal phase. 
For $x>x_c$ one can find the coexistence curve in $(x, n)$ space. Suppose that 
a volume fraction $v$ is occupied by the $NP$ phase and the rest ($1-v$) by the 
superfluid phase. 
Then:
\begin{equation}
n(v) =vn_{NP} + (1-v)n_{SF}\\
\end{equation}

\begin{equation}
x(v) =\frac{(\frac{x_c}{1+x_c}n_{NP}
  +(1-v)\frac{n_{SF}}{2})}{(\frac{1}{1+x_c}n_{NP}+(1-v)\frac{n_{SF}}{2})}\;,
  ~\label{eq:coexcurve}
\end{equation}
where $v\in[0, 1]$.

The parametric dependence on $v$ can be eliminated to find $g(x, n_{NP})$ along the
coexistence curve. The coexistence curve (for example, see dashed curve in
Fig.~\ref{fig:coex}) should not be seen as the continuation of the $NP$ curve
at constant density, since the density changes along it. It is simply a
projection of the coexistence curve in $(g, x, n)$ space, on the $(g, x)$
plane.

To proceed with the calculation of the coexistence curve, we need to calculate 
$g(x, n)$ for the $NP$ phase and $g_{SF}(n)$ for the $SF$ phase. We use Quantum
Monte Carlo techniques to calculate the energies at a few values of $(x, n)$
and interpolate to determine the functional dependence. This is discussed next.

\section{Green's Function Monte Carlo simulations}
\label{section:MonteCarlo}

The Hamiltonian for low-density neutron matter is:
\begin{equation}
{\cal{H}} = \sum\limits_{k = 1}^{N}  ( - \frac{\hbar^2}{2m}\nabla_k^{2} )  + \sum\limits_{i<j'} v(r_{ij'})~.
\end{equation}
where $N$ is the total number of particles. 
The neutron-neutron interaction is not purely $s$-wave but still somewhat simple if one considers
the AV4' formulation: \cite{Wiringa:2002}
\begin{equation}
v_4(r) = v_c(r) + v_{\sigma}(r){\mbox{\boldmath$\sigma$}}_1\cdot{\mbox{\boldmath$\sigma$}}_2,
\label{vfour}
\end{equation}
In the case of $S$=0 (singlet) pairs this gives:
\begin{equation}
v_S(r) = v_c(r) - 3v_{\sigma}(r)~.
\label{ves}
\end{equation}
However, it also implies an interaction for $S$=1 (triplet) pairs:
\begin{equation}
v_P(r) = v_c(r) + v_{\sigma}(r)~.
\end{equation}
Ref. \cite{Gezerlis:2010}  explicitly included such $p$-wave interactions in 
the same-spin pairs (the contribution of which was small even at the highest density considered), 
and perturbatively corrected the $S=1, M_S=0$
pairs to the correct $p$-wave interaction.

In these calculations it is customary to first employ a 
standard Variational Monte Carlo simulation, which
minimizes the expectation value of the Hamiltonian
given a variational wave function $\Psi_V$.
At a second stage, the output of the Variational Monte Carlo calculation
is used as input in a fixed-node Green's Function Monte Carlo simulation,
which  
projects out the lowest-energy eigenstate
$\Psi_{0}$ from the trial (variational) wave function $\Psi_{V}$. This is 
accomplished
by treating the Schr\"{o}dinger equation as a diffusion equation in imaginary
time $\tau$ and evolving the variational wave function up to large $\tau$.
The ground state is evaluated from:
\begin{eqnarray}
\Psi_0 & = & \exp [ - ( H - E_T ) \tau ] \Psi_V  \\ \nonumber
& = & \prod \exp [ - ( H - E_T ) \Delta \tau ] \Psi_V,
\end{eqnarray}
evaluated as a branching random walk.
The fixed-node approximation gives a wave function $\Psi_0$ that is the
lowest-energy state with the sames nodes (surface where $\Psi$ = 0) as the
trial state $\Psi_V$. The resulting energy $E_0$ is an upper bound to the
true ground-state energy.  

The ground-state energy $E_0$ can be obtained from:
\begin{equation}
E_0 = \frac{ \langle \Psi_V | H | \Psi_0 \rangle}{ \langle \Psi_V | \Psi_0 \rangle}
= \frac{ \langle \Psi_0 | H | \Psi_0 \rangle}{ \langle \Psi_0 | \Psi_0 \rangle}.
\end{equation}

The variational wave function is taken to be of the following form:
\begin{equation}
\Psi_V({\bf{R}}) =  \prod\limits_{i \neq j} f_P(r_{ij}) \prod\limits_{i' \neq j'} f_P(r_{i'j'}) \prod\limits_{i,j'} f(r_{ij'}) \Phi({\bf{R}})~.
\label{varwave}
\end{equation}
In our earlier works we studied superfluid neutron matter and therefore
used a trial wave function of the Jastrow-BCS form with fixed
particle number. \cite{Gezerlis:2008,Gezerlis:2010,Gezerlis:2011}
In the present work we are attempting to determine the stability 
of different phases. In order to do this, we have completely mapped
out the energy of a normal (i.e. non-superfluid) neutron gas as a
function of the polarization at different densities. Thus, 
the $\Phi({\bf{R}})$ we used in Eq. (\ref{varwave}) 
describes the particles as being in a free Fermi gas 
(i.e., lets all the correlations lie within the Jastrow functions). 
In other words, the wave function is composed of two Slater determinants
(one for spin-up particles and one for spin-down ones):
\begin{equation}
\Phi_S({\bf{R}}) = \Phi_S({\bf{R}})_{N_{\uparrow}} \Phi_S({\bf{R}})_{N_{\downarrow}'}
\label{slater}
\end{equation}
where 
\begin{equation}
\Phi_S({\bf{R}})_{N_{\uparrow}} = {{\cal A}}[\phi_n(r_1) \phi_n(r_2) \ldots \phi_n(r_{N_{\uparrow}})]
\end{equation}
and
\begin{equation}
\Phi_S({\bf{R}})_{N_{\downarrow}'} = {{\cal A}}[\phi_n(r_{1'}) \phi_n(r_{2'}) \ldots \phi_n(r_{N_{\downarrow}'})]
\end{equation}
The primed (unprimed) indices correspond to
spin-up (spin-down) neutrons and $N_{\uparrow} + N_{\downarrow}' = N$. 
${\cal A}$ is the antisymmetrizer and 
$\phi_n(r_k) = e^{ i {\bf{k}}_{\bf n} \cdot {\bf r}_k}/L^{3/2}$. 

The Jastrow part is usually taken from a
lowest-order-constrained-variational method
calculation described by a Schr\"{o}dinger-like equation:
\begin{equation}
- \frac{\hbar^2}{m}\nabla^{2} f(r)  + v(r) f(r) = \lambda f(r)~\nonumber
\end{equation}
for the opposite-spin $f(r)$ and by a corresponding equation for the
same-spin $f_P(r)$. Since the $f(r)$ and $f_P(r)$ are nodeless, they do
not affect the final result apart from reducing the statistical error. 

We calculate ground-state energies at different total number densities [$n =
(N_{\uparrow} + N_{\downarrow})/L^3$], more specifically at $n_1 = 6.65 \times
10^{-4}$, $n_2 = 2.16 \times 10^{-3}$, and $n_3 = 5.32 \times 10^{-3}$
fm$^{-3}$.  To put these densities into perspective we can compare them to
nuclear matter saturation density: they are $0.41, 1.35$, and $3.32$ percent,
respectively, of $n_0 = 0.16$ fm$^{-3}$.  These simulations were performed at
values of the relative fractions chosen specifically in order to ensure a full
coverage of the $x$-axis. More specifically, simulations were carried out at
relative fractions of $x={{0},\; {0.333},\; {0.579},\; {0.818},\; {1}}$,
corresponding to particle numbers of ${{33+0},\; {57+19},\; {57+33},\;
{33+27},\; {33+33}}$, respectively. These systems are quite large (and
therefore computationally demanding) so as to ensure a minimization of
finite-size effects (see also Refs. \cite{Gezerlis:2008,
Gezerlis:2010,Forbes:2011a}).

\begin{figure}[t]
\vspace{0.5cm}
\begin{center}
\includegraphics[width=0.46\textwidth]{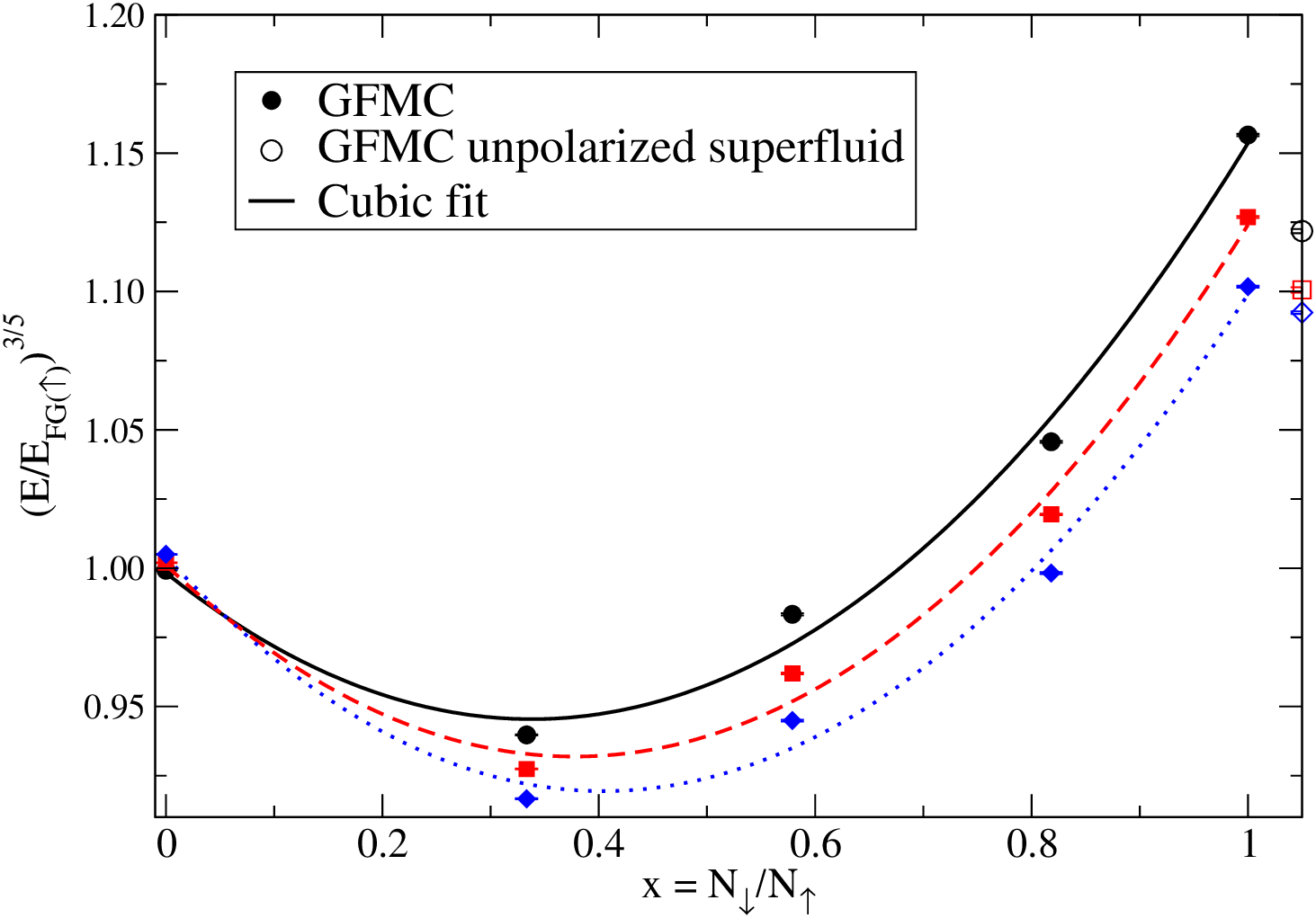}
\caption{(color online) 
Ground-state energy per particle (in units of the free Fermi gas energy)
scaled to the power of $3/5$ for normal spin-polarized neutron matter. 
Shown are QMC results at three different total densities $n_1, n_2, n_3$ 
(as points: black circles, red squares, blue diamonds, respectively), 
along with cubic fits to the Monte Carlo results 
(as lines: black solid, red dashed, blue dotted, respectively).
Also shown using hollow symbols are the results for an unpolarized superfluid.}
\label{fig:ener}
\end{center}
\end{figure}

The results are shown in Fig.~\ref{fig:ener} as points (circles, squares, and diamonds, 
respectively), along with cubic fits to the microscopic data. The latter will be useful
to us in the following section, when we try to check the relative stability 
of different phases. To facilitate the use of these results in connection
with Eq. (\ref{eq:gintro}), we have divided the ground-state energies with the 
energies of corresponding free spin-up Fermi gases, and raised the result
to the 3/5 power. When $x=0$ we see that the values are very close to 1, 
though still above it: this is due to the same-spin $p$-wave interactions.
Similarly, as we increase the density the energy is decreased, a fact
that, as we see in Fig.~\ref{fig:ener}, holds at any relative fraction $x$.
The statistical errors of the Quantum Monte Carlo results are smaller than 
the symbols shown in the figure.
Also shown to the right of the figure are the QMC results for an upolarized
superfluid at the three densities of interest (hollow circles, squares, and diamonds, 
respectively). The results for $n_1$ and $n_2$ were taken from Ref.~\cite{Gezerlis:2010};
the value at $n_3$ was re-optimized for the purposes of the present work.

\section{Phase competition}
\label{section:PhaseCompetition}

\begin{figure}[t]
\vspace{0.5cm}
\begin{center}
\includegraphics[width=0.46\textwidth]{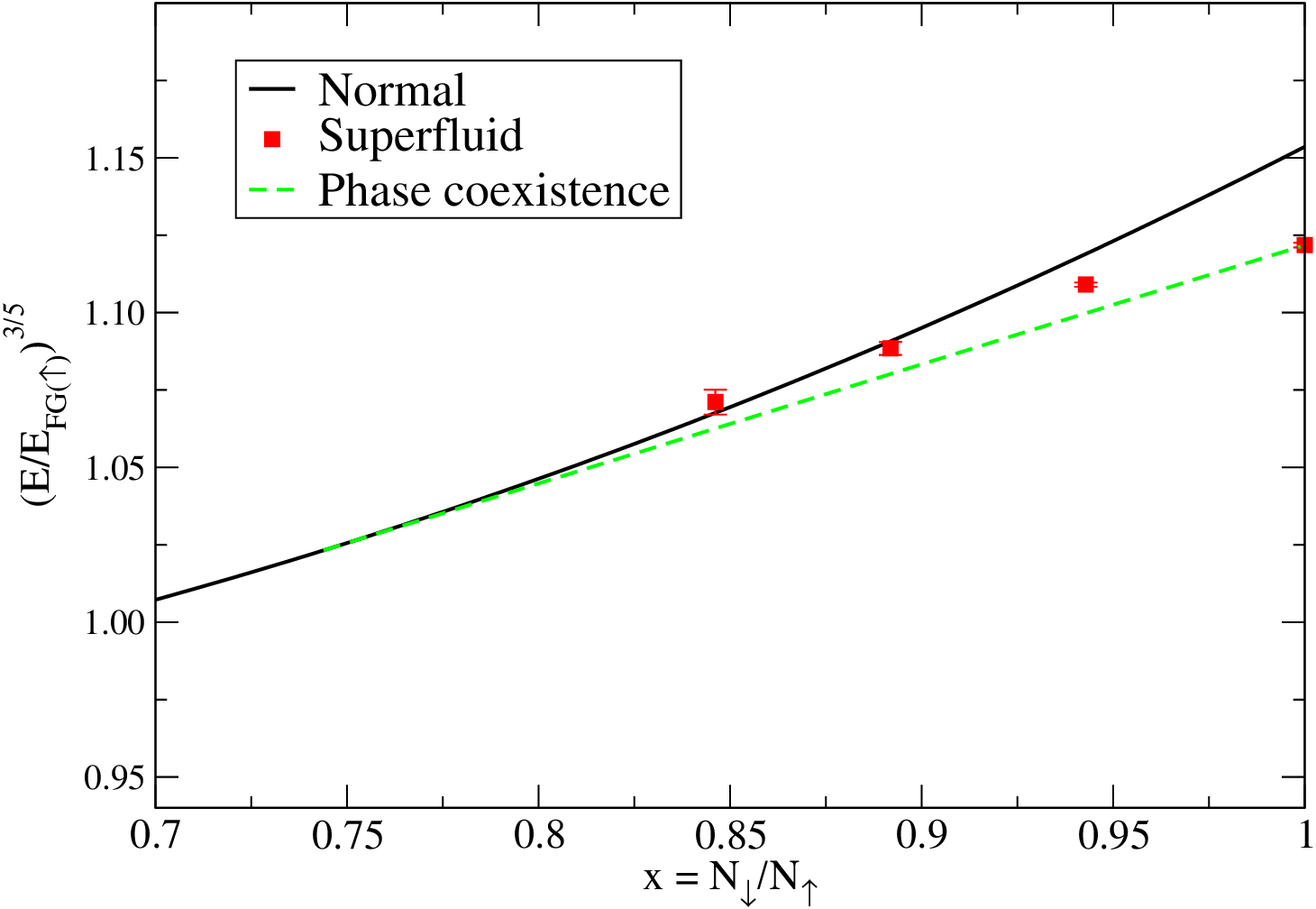}
\caption{(color online) 
The construction of the mixed phase at density $n_1$.
Shown is the energy of a normal neutron gas from Fig.~\ref{fig:ener}
of the present work (black solid line), along with the homogeneous
superfluid results of Refs. \cite{Gezerlis:2010,Gezerlis:2011} (red
square points), and a tangent construction showing the phase
coexistence curve (green dashed line). For large values of the relative fraction,
the phase separated state is energetically favored with respect
to a homogeneous polarized superfluid.}
\label{fig:coex}
\end{center}
\end{figure}

Having arrived at the functional dependence of the energy of a normal neutron
gas on the relative fraction (thereby getting information on the $g(x,n)$
function defined in Eq.~(\ref{eq:gintro}) and appearing in the rest of
Sec.~\ref{section:MixedPhase}), we can now use the simple cubic fits for the
three densities $n_1, n_2, n_3$ to determine the competition between
different phases. More specifically, Fig.~\ref{fig:ener} shows  the behavior of
the NP (normal polarized) equation of state.  From the hollow points on the same figure we
also have access to the energy of the SF (unpolarized superfluid) phase. Using
these two dependences, we can find out the critical values of the relative
fraction ($x_{c1}, x_{c2}, x_{c3}$, respectively) at the three densities we are
studying. Above a critical relative fraction the system phase separates into
a mixture of SF and NP, whereas below that value the neutron gas is normal, and
therefore does not exhibit any of the widely studied features of a superfluid
in the neutron star crust. 

To determine $x_c$ we need the functional dependence of $g(x,n)$ on both $x$
and $n$, which is given by the interpolations for $n_1$, $n_2$ and $n_3$.
Let us first focus on $n_2$. In Eq.~(\ref{eq:mu}) we also
need the derivative with respect to $n$. We estimate this derivative by finite
differences: $(g(x, n_2)-g(x, n_1))/(n_2-n_1)$ and 
$(g(x, n_3)-g(x, n_2))/(n_3-n_2)$. We estimate the derivatives for
$SF$ used in Eq.~\ref{eq:muSF} with a linear form in $1/(k_Fa)$ and $k_Fr_e$.
In practice, $\frac{\partial g}{\partial n}$ is small and doesn't
affect the value of $x_c$ and $n_{SF}$ significantly. We can then solve the
conditions for coexistence and find $x_c$ and $n_{SF}$.
The numerical values show slight dependence on the interpolation method used.
A simple cubic fit as a function of $x$ gives $x_{c2}=0.78$ and
$n_{SF}= 2.17 \times 10^{-3}$ fm$^{-3}$.  A three parameter fit in $(1, x,
x^{5/3})$ motivated by~\cite{Giorgini:2008,Lobo:2006} gives $x_{c2}=0.77$ and
$n_{SF}= 2.18 \times 10^{-3}$ fm$^{-3}$.  This gives some idea about the
errors made in the interpolation: we estimate our error bar on $x_c$ to be
$0.02$ for the three densities. This detailed procedure gives results very
similar to a simpler procedure, namely constructing a tangent to the constant
density $NP$ curve for $g(x, n)$ passing through the value of $g_{SF}(n)$ at
the same density, which gives $x_{c2} = 0.78$. The close agreement between the
two approaches stems from the fact that $\frac{\partial g}{\partial n}$ is
small --- for $\frac{\partial g}{\partial n}=0$, the coexistence curve is given
exactly by the tangent construction~\cite{Bulgac:2007}. Therefore, we will use
the tangent construction to find $x_c$ for $n_1$ and $n_3$ below.

This analysis requires that for $x<x_c$ the normal phase is thermodynamically
stable. That is, the eigenvalues of the matrix $\frac{d^2 {\cal{E}}}{dn_i
dn_j}$ are positive. For $n=n_2$ we can check and have checked explicitly that
this is the case. Finally, we find $\delta\mu$ at $x_{c2}$ is given by
$\delta\mu_{c2}/E_{F}=0.16(2)$ where $E_{F}=({k_F^2}/{2m})$.  Comparing with
$\Delta_{SF2}/E_{F}\sim0.23$ obtained from Ref.~\cite{Gezerlis:2010}, we
see that $\Delta_{SF2}>\delta\mu_{c2}$ satisfying the third condition for
coexistence.

In Fig.~\ref{fig:coex} we have plotted as a black solid line the $n_1$ results
from Fig.~\ref{fig:ener}; at this density the perturbative $S=1, M_S=0$
correction in the Hamiltonian mentioned in Sec.~\ref{section:MonteCarlo}
is the smallest, leading to the highest degree of confidence in the accuracy of
our microscopic results.  We focus on large values on the $x$-axis for
reasons that will soon become clear.  We also show as red square points the
unpolarized superfluid result from Ref.~\cite{Gezerlis:2010} (at $x=1$) and the
homogeneous polarized superfluid results of Ref. \cite{Gezerlis:2011} (at
smaller $x$). In addition, we show a coexistence curve which is arrived at by a
tangent construction from the unpolarized superfluid to the normal polarized
curve when the $y$-axis is chosen as we have~\cite{Bulgac:2007}. This
coexistence line meets the NP curve at $x_{c1} = 0.74$
($\delta\mu_{c1}/E_{F}=0.2$). As can be clearly seen
in this Figure, the coexistence line lies below the homogeneous polarized
superfluid Quantum Monte Carlo results, implying that these are not
energetically favored (foreseeing this possibility, Ref. \cite{Gezerlis:2011}
included a critical assumption explicitly excluding the
possibility of phase separation). This situation follows the behavior
of ultracold fermionic gases at unitarity: \cite{Lobo:2006, Pilati:2008,
Giorgini:2008} there the critical value of the relative fraction was closer to
$x_c = 0.44$. The value we find is different for a variety of reasons: first,
$n_1$ corresponds to $k_F a = -5$, which is somewhat smaller than $k_F a =
\infty$. More importantly, the neutron-neutron interaction is characterized by
a finite effective range and the presence of higher partial waves. The latter
point explains why when we repeat this exercise for $n_2$ ($k_F a = -7.4$) and
$n_3$ ($k_F a = -10$) the corresponding results (from tangent constructions)
for the critical fraction are $x_{c2} = 0.78$
($\delta\mu_{c2}/E_{F}=0.16$) and $x_{c3} = 0.88$
($\delta\mu_{c3}/E_{F}=0.07$), respectively.  The trend appears to be
that at larger density (stronger coupling) the critical fraction is moving
toward 1, implying that farther down toward the core of the neutron star
polarized neutron matter quickly becomes normal for the vast majority of
possible polarizations.

\begin{figure}[t]
\vspace{0.5cm}
\begin{center}
\includegraphics[width=0.46\textwidth]{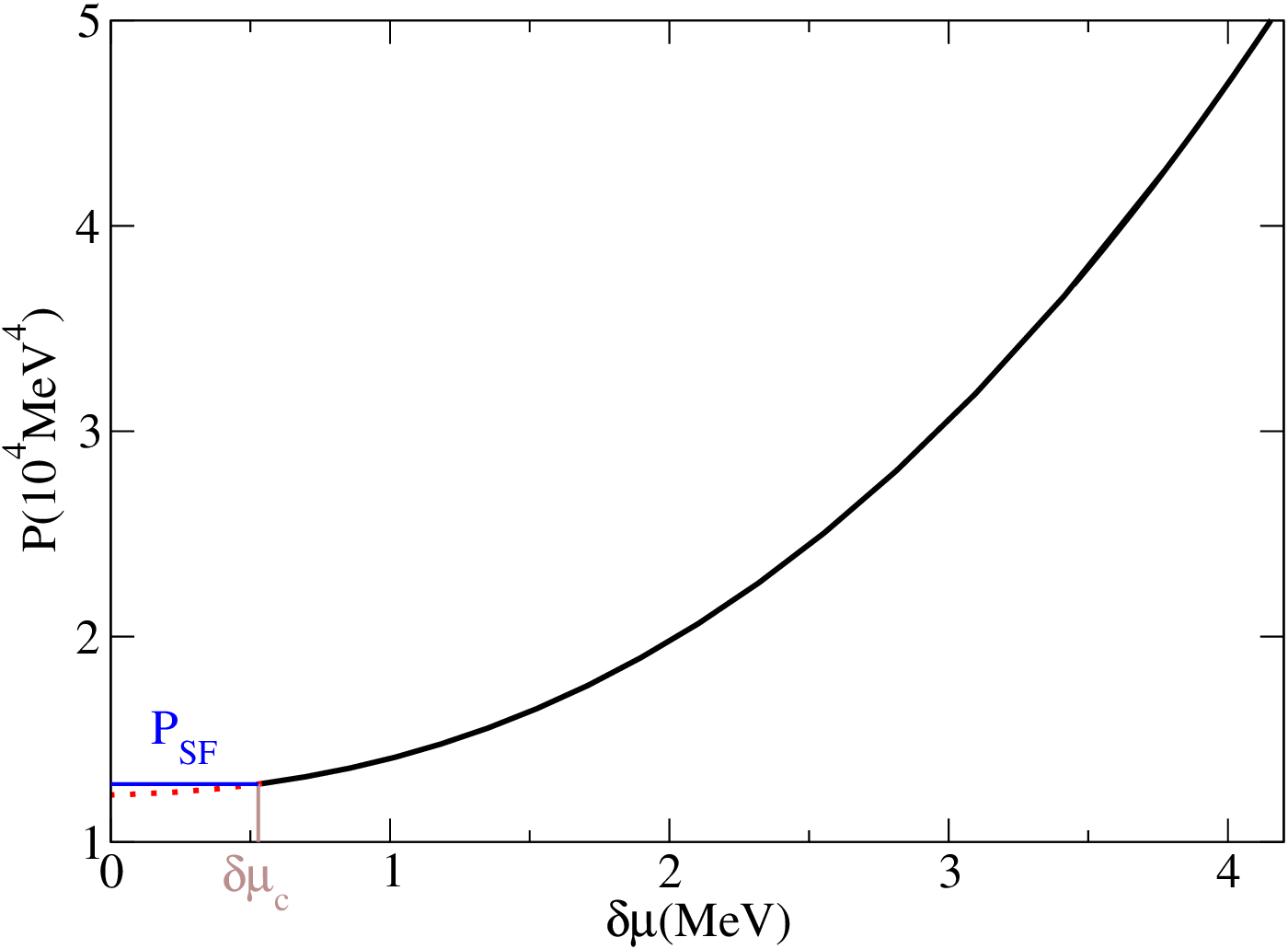}
\caption{(color online) Pressure as a function of $\delta\mu$ for fixed
$\mu=1.94$ MeV. For $\delta\mu>\delta\mu_c=0.53$ MeV the unpaired phase (bold
black curve) has a higher pressure. At $\delta\mu_c$ [marked by the vertical line (brown online)], the $SF$
and the $NP$ phases can coexist. For smaller $\delta\mu$, the pressure
of the $SF$ phase marked by the horizontal line (blue online) is larger
than the pressure of the $NP$ phase marked by the dashed line (red
online), and the $SF$ phase is favored. On
the other extreme for $\delta\mu=4.18$ MeV, $x=0.0005$. We choose $\mu$ so that the
density of the normal phase at $\delta\mu_c$ is $n_2$.}
\label{fig:GrandCan}
\end{center}
\end{figure}

The value of $\delta\mu_c$ is relevant for neutron star phenomenology since
whether $\delta\mu$ is greater or less than $\delta\mu_c$ determines the phase. 
This is seen more simply in the grand canonical ensemble where we can compare
the pressure as a function of the chemical potentials. In Fig.~\ref{fig:GrandCan} we plot
the pressure as a function of $\delta\mu$ for fixed $\mu=1.94$ MeV. For
this value of $\mu$ the density of the superfluid phase at the critical
$\delta\mu$ is $n_2$. For $\delta\mu<\delta\mu_c$ the pressure of the $SF$
phase is larger and the system is unpolarised. Even if the magnetic field is
not large enough to polarize the system, certain thermal observables, for
example the specific heat, are sensitive to the difference between $\delta\mu$
and $\delta\mu_c$.

Determining the critical relative fractions is also relevant: if a magnetic
field is strong enough to polarize neutron matter sufficiently ($x\leq x_c$),
the gas ceases to be superfluid even at zero temperature.  At such large
magnetic fields, processes that rely on the presence of a bulk superfluid in
the inner core of neutron stars, for example a recently proposed
heat-conduction mechanism which requires superfluid phonons
\cite{Aguilera:2009}, would no longer be operational.  On a different note, a
precise determination of these fractions (along with the rest of the equation
of state) is also significant for Skyrme functional practitioners: for the
region where the NP phase is the true ground state of the system, the results
shown in Fig.~\ref{fig:ener} are a microscopic constraint to nuclear
energy-density functional theory, which can help improve its predictions on
neutron-rich heavy nuclei.

\section{Summary \& Conclusions}

In summary, we have studied spin-polarized low-density neutron matter at 
many values of the polarization. Our Quantum Monte Carlo approach 
provides tight upper bounds which are expected to be quite close to the
true ground state-energy of the system. At very-low density
the Hamiltonian becomes quite simple and we include its 
dominant well-known terms. 
We have calculated the energy as a function of polarization using the 
AV4$'$ interaction at three different densities. 

One of our main results is a determination of the critical relative
fraction values above which the system phase separates into an unpolarized
superfluid and a polarized normal neutron gas. Above these values, a 
homogeneous polarized superfluid is found not to be energetically favored.
Below these values, the neutron gas is completely normal. Both these facts
can lead to astronomically important consequences. Importantly, the trend of our 
results seems to imply that at slightly larger densities
$x_c$ would be very close to one: this would mean that even at zero
temperature a small polarization would be enough to close
the superfluid pairing gap, and thus easily lead to larger polarization.
In a non-astrophysical context, it is conceivable that our
results could be tested directly by using ultracold fermionic atom gases with unequal
spin populations and a finite effective range. In cold atoms, 
Quantum Monte Carlo simulations of spin-polarized matter
have in recent times been repeatedly used as input to 
computationally less demanding density-functional theory approaches. 
Similarly, we expect that the results presented in this work 
can also be used as input to self-consistent mean-field models of nuclei.

A possible direction for future work, that would be interesting to both nuclear astrophysicists
and Skyrme practitioners, would be a study of the behavior of polarized and
unpolarized superfluid and normal neutron matter in an external periodic potential.
Such a study of the static response of neutron matter could also in principle
be guided (or guide) analogous research related to optical lattice experiments
with cold atoms.

In the neutron star context, it will be useful to analyze the finite
temperature phase diagram to understand how the specific heat is modified as a
function of the chemical potential splitting, and how this modification affects
transient phenomena in the neutron star crust.

\begin{acknowledgments}
A.G. thanks George Bertsch, Kai Hebeler, and Achim Schwenk for
stimulating discussions. R.S. acknowledges discussions with Jeremy
Heyl.  The authors thank Mark Alford and Sanjay Reddy for valuable comments and
suggestions. The authors acknowledge the hospitality of the Perimeter Institute
and the organizers of the Micra 2011 workshop where this work was initiated.
This work was supported by DOE Grant No.  DE-FG02-97ER41014 and the National
Sciences and Engineering Research Council of Canada (NSERC).  Computations were
performed at the US DoE's National Energy Research Scientific Computing Center
(NERSC). 
\end{acknowledgments}

%%%%%%%%%%%%%%%%%%%%%%%%%%%%%%%%%%%%%%%%%%%%%%%

\end{document}